\documentclass[twocolumn,10pt]{article}
\usepackage{usenix}
\usepackage{times}

\usepackage{balance}
\usepackage{bibspacing}
\usepackage[compact]{titlesec}

\DeclareMathAlphabet{\mathcal}{OMS}{cmsy}{m}{n}

\usepackage[pdftex]{graphicx}
\graphicspath{{graphs/}{images/}}
\DeclareGraphicsExtensions{.pdf}

\usepackage[cmex10]{amsmath}
\usepackage[font=small]{caption}
\usepackage[caption=false,font=small]{subfig}
\usepackage{xspace}

\usepackage{rotating}
\usepackage{multirow}
\usepackage{multicol}

\usepackage{enumitem}

\usepackage[hyphens]{url}
\usepackage[usenames,dvipsnames]{xcolor}
\usepackage[colorlinks=true, citecolor=OliveGreen, linkcolor=BrickRed, urlcolor=MidnightBlue, final, pdftex]{hyperref}

\usepackage{cite} 
\usepackage{algorithm}
\usepackage{algorithmic}

\usepackage{fancyhdr}
\usepackage{tikz}
\newcommand*\circled[1]{\tikz[baseline={([yshift=-.8ex]current bounding box.center)}]{\node[shape=circle,draw,inner sep=.1pt,fill=black] (char) {\color{white}{\scriptsize{\textbf{#1}}}};}}

\newcommand{\etal}{{\em et~al.}~}

\renewcommand{\paragraph}[1]{\noindent\textbf{#1:}}
\renewcommand{\S}{Section~}

\newcommand{\ignore}[1]{}

\newcommand{\thetitle}{Tor's Been KIST: A Case Study of Transitioning Tor Research to Practice}

\hypersetup
{
    pdfauthor={Rob Jansen and Matthew Traudt},
    pdfsubject={Anonymous Communication Performance},
    pdftitle={\thetitle},
    pdfkeywords={Anonymous Communication, Anonymity, Tor, Simulation, Performance, Deployment}
}

\title{\thetitle}

\author{
{\rm Rob Jansen and Matthew Traudt}\\
U.S. Naval Research Laboratory\\
\{rob.g.jansen, matthew.traudt\}@nrl.navy.mil
}

\date{}

\begin{document}

\maketitle


\begin{abstract}
Most computer science research is aimed at solving difficult problems with a
goal of sharing the developed solutions with the greater research community. For
many researchers, a project ends when the paper is published even though a much
broader impact could be achieved by spending additional effort to transition
that research to real world usage. In this paper, we examine the opportunities
and challenges in transitioning Tor research through a case study of deploying a
previously proposed application layer socket scheduling policy called KIST into
the Tor network. We implement KIST, simulate it in a 2,000-relay private Tor
network using Shadow, deploy it on a Tor relay running in the public Tor
network, and measure its performance impact. Confirming the results reported in
prior research, we find that KIST reduces kernel outbound queuing times for
relays and download times for low-volume or bursty clients. We also find that
client and relay performance with KIST increases as network load and packet loss rates
increase, although the effects of packet loss on KIST were overlooked in past
work. Our implementation will be released as open-source software for inclusion
in a future Tor release.
\end{abstract}

\section{Introduction} \label{sec:intro}
\vspace{-1mm}
Tor~\cite{torproject,tor-design} is the most popular system for anonymous communication,
consisting of roughly 7,500 volunteer-operated
\textit{relays} 
transferring over 100 Gbit of traffic every
second~\cite{metrics}. Tor has roughly two million unique daily
users~\cite{metrics}, over 500,000 of which use Tor at any given
time~\cite{privcount-ccs2016}. Clients using Tor construct \textit{circuits} of
three Tor relays through which they tunnel their Internet connections, while
a \textit{hidden onion service} protocol allows both clients and
servers to create circuits and connect them inside of the Tor network in order for
 both endpoints to achieve end-to-end encryption and anonymity. 

Tor is designed to provide low-latency anonymity: relays immediately forward
packets without introducing any artificial delays in order to provide a usable
experience for clients that use Tor to browse Internet websites. However, Tor's
three relay-hop design (six relay-hops for hidden onion services) combined with its
popularity and available resources results in significantly longer transfer
times compared to direct connections. 

There has been a significant amount of research into performance enhancements
for the Tor network~\cite{psitor2016}, including proposals that change the way
Tor relays classify~\cite{jansen2012throttling,ccs2012-classification} and
prioritize~\cite{tang2010scheduling,jansen2014kist} traffic
and handle relay connections~\cite{wpes12-torchestra,nowlanreducing,ccs2013pctcp,wpes2014imux,usec2009dtls,jansen2014kist}.
Relays currently use the circuit scheduling approach of Tang and
Goldberg~\cite{tang2010scheduling}: an
exponentially-weighted moving average (EWMA) of the throughput of each circuit
is used to prioritize low-volume, bursty traffic over high-volume, bulk traffic.
Jansen \etal identified flaws in the way that relays write data
to the kernel that were significantly reducing the effectiveness of the intended
priority mechanisms, and proposed a new socket scheduling policy called 
Kernel-Informed Socket Transport (KIST) to overcome these challenges~\cite{jansen2014kist}.
%
%
A KIST prototype was evaluated in Tor network simulations using
Shadow~\cite{jansen2012shadow,shadowdev} and was shown to reduce kernel outbound queuing
times and end-to-end latency for low-volume traffic; however, it was never tested or
evaluated on any relays running in the live, public Tor network.

\vspace{-1mm}
In this paper, we present our work in further understanding the impact that KIST
has on client and relay performance, with the goal of producing a
production-level implementation of KIST that is suitable to include in a future
Tor release. Toward this goal, we first independently implement
KIST in collaboration with the Tor developers: we discuss the details of this implementation and the supporting architecture in \S\ref{sec:deploy}.
Our KIST implementation improves
upon the previous prototype~\cite{jansen2014kist} by significantly reducing the
overhead involved in managing the process of writing to sockets, and contains
the components that would be required for our code to be included in Tor (e.g.,
unit tests and documentation).

\vspace{-1mm}
We then simulate KIST in a large scale private Tor network of 2,000 relays and
up to 70,000 clients using Shadow~\cite{jansen2012shadow,shadowdev}, both to
test our code and to confirm past research~\cite{jansen2014kist} reporting that
KIST is capable of improving performance for Tor relays and clients. Our results
in \S\ref{sec:simeval} confirm the results from prior work: KIST is capable of
relocating congestion from the kernel output queue into Tor where it can be
better managed and priority can be properly applied. Additionally, the
effects of packet loss on KIST were not considered in prior research; we extend that
research by analyzing KIST under a range of network load and packet loss models.
We find that KIST performs at least as well as the default Tor scheduler across
all tested conditions, and that KIST is able to increasingly improve both client
and relay performance relative to Tor's default scheduler as both network load
and packet loss rates increase. We provide the first indication that KIST
effectively backs off of high-volume circuits under high loss while correctly
prioritizing low-volume or bursty circuits.

We also provide the first live-network evaluation of KIST in
\S\ref{sec:neteval}, using a deployment of KIST on a fast relay in the public
Tor network. We find that Tor application queuing time increases with KIST as
expected; however, we are unable to detect a significant change in relay
throughput or kernel outbound queuing time that can be attributed to KIST. We
believe that this is partially due to our lack of experimental control over Tor
network effects, and partially because our relay did not experience enough load
or packet loss for KIST to significantly influence the socket scheduling
process. We also find that KIST overhead is tolerable: with our optimizations
and suggested parameter settings, the system call overhead scales linearly with
the number of relay-to-relay TCP connections with write-pending data, and
independently of the total number of open sockets.

Finally, we briefly discuss the lessons that we learned while producing and
evaluating deployable Tor code, and generalize our experiences to provide
insights into the process of transitioning Tor research.


\section{Background} \label{sec:background}

In this section we provide background on Tor and how it handles traffic, and
describe how KIST changes Tor's scheduling mechanisms.

\subsection{Tor} \label{sec:background:tor}

Tor~\cite{tor-design} is a low-latency anonymity network that is primarily used
to access and download webpages and to transfer files~\cite{privcount-ccs2016},
but can be used to facilitate anonymous communication between any pair of
communicating peers in general.

To use Tor, a client first constructs a \textit{circuit} by telescoping an
encrypted connection through an \textit{entry} relay, a \textit{middle} relay,
and an \textit{exit} relay, and then requests the exit relay to connect to the
desired external destinations on the client's behalf. The logical connections
made by clients to destinations are called \textit{streams}, and they are
multiplexed over circuits according to the policy set by each circuit's
exit relay. The client and the exit relay package all application-layer payloads
into 512-byte \textit{cells}, which are onion-encrypted and forwarded through
the circuit.
When a client and a server communicate using the onion service protocol, they
both construct circuits and connect them at a client-chosen rendezvous relay;
the resulting six relay-hop circuit is then used to provide end-to-end
encryption.

\begin{figure}[t]
  \centering
  \includegraphics[width=\linewidth]{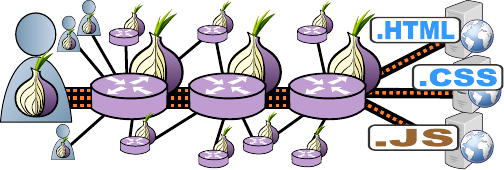}
  \caption{Connections between Tor clients, relays, and destinations. Solid black lines represent TCP connections, and dashed orange lines represent logical streams of client application payloads that are multiplexed over both the TCP connections and the encrypted circuits that are constructed by clients. High performance relays generally maintain thousands of open network sockets as a result of this architecture.}
  \label{fig:conns}
\end{figure}

Circuits are multiplexed over TCP connections which are maintained between Tor
clients, relays, and destinations (see Figure~\ref{fig:conns}).
During the circuit construction process, each relay will create a TCP connection
to the next-hop relay chosen by the client if such a connection does not already
exist. Although idle TCP connections are closed
to conserve resources, each relay may maintain up to $n - 1$ open TCP
connections to other relays for onion routing in a network consisting of $n$ relays. In addition
to the relay TCP connections, entry relays maintain TCP connections with clients
while exit relays initiate and maintain TCP connections with destination servers,
e.g., to download webpages and embedded objects. Therefore, high-bandwidth
relays generally maintain thousands of open network sockets at any given time.

\subsection{Tor Traffic Management} \label{sec:background:traffic}

We now describe how a Tor relay internally handles and forwards traffic
prior to version \texttt{0.2.6.2-alpha}, i.e.,
before merging support for KIST. We describe how this architecture
was modified to support KIST in newer versions of Tor in \S\ref{sec:deploy}.

\begin{figure}[t]
  \centering
  \includegraphics[width=\linewidth]{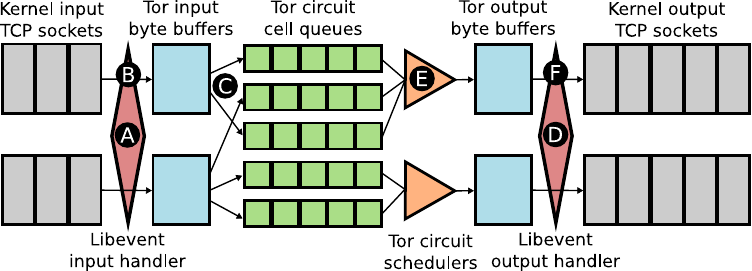}
  \vspace{-5mm}
  \caption{Data is transferred through Tor using several layers of buffers and queues. The transfer process is driven by libevent, an event notification library. Circuit schedulers attempt to prioritize low-volume or bursty traffic (web browsing) over high-volume, bulk traffic (file sharing).}
  \label{fig:buffers}
\end{figure}

Tor's traffic handling process involves several layers of buffers and
schedulers, and is driven by an event notification library called
libevent\footnote{\url{http://libevent.org}} (see Figure~\ref{fig:buffers}). Tor registers new
sockets with libevent and uses it to asynchronously poll those sockets in order
to track when they are readable and writable. There is an input and output byte
buffer corresponding to each socket that is used to buffer kernel reads and
writes, respectively. There is also a circuit scheduler corresponding to each
socket that is used to prioritize traffic, as we will further describe below.

When a TCP socket is readable, i.e., has incoming bytes that can be read from
the kernel, libevent notifies Tor by executing a
callback function (Figure~\ref{fig:buffers}\circled{A}). Tor then reads input bytes
from the readable TCP socket into the input byte buffer using OpenSSL, which
removes the transport layer of encryption (Figure~\ref{fig:buffers}\circled{B}). For each
512-byte chunk that accumulates in the input buffer, Tor
creates a cell, applies onion
encryption to it, and then either handles the cell directly if
possible or moves it to the circuit queue corresponding to its next-hop relay
(Figure~\ref{fig:buffers}\circled{C}). The cells remain in the circuit queue until
they can be written to the outgoing TCP connections.

When a TCP socket is writable, i.e., has available space such that outgoing
bytes can be written to the kernel, libevent notifies Tor by executing a
callback function (Figure~\ref{fig:buffers}\circled{D}). 
Because circuits are multiplexed over TCP connections, there may be several
circuits with cells that are pending to be written to the writable socket; there
is a circuit scheduler corresponding to each socket that is used to determine
the order that pending cells are written.
The circuit scheduler corresponding to the writable socket is invoked to choose the
circuit with the best priority and write a cell from it to the output byte buffer
(Figure~\ref{fig:buffers}\circled{E}). Tor uses the circuit
scheduling algorithm of Tang and Goldberg~\cite{tang2010scheduling} to determine
circuit priority. Their algorithm is based on an exponentially-weighted moving
average (EWMA) of circuit throughput and prioritizes circuits carrying
low-volume or bursty traffic over those carrying high-volume, bulk traffic.
Bytes from the output buffer are written to the kernel using OpenSSL
(Figure~\ref{fig:buffers}\circled{F}) when the output buffer length exceeds a
threshold of 32 KiB and when circuit scheduling for the writable socket is
complete.

Note that the entire reading and writing processes are driven by libevent, which
issues readable and writable notifications for one socket at a time.
The order in which these notifications are delivered is not configurable.

\subsection{KIST}

In prior work, Jansen \etal observed two major problems with Tor's traffic management
process, and designed a new socket scheduler, called
kernel-informed socket transport (KIST), to correct these issues~\cite{jansen2014kist}.

\paragraph{Problem: Sequential Socket Writes}
Because Tor only considers a single socket out of potentially thousands or tens
of thousands of sockets at a time, it is possible that the circuit priority
scheduler would only consider a small fraction of the circuits that could have
been written at any given time. They showed through experimentation that circuit
priority mechanisms are ineffective when multiple circuits of different priority
levels do not share the same outgoing socket (since then they are not considered
by the same circuit scheduler instance). Because of Tor's sequential, single-socket output
handler, worse priority high-volume traffic from one socket would be written
to the kernel before better priority low-volume traffic from another socket.
Additionally, the first-in-first-out kernel scheduler would send the
already-written worse-priority data before any better-priority data that may
arrive and be written an instant later.

\paragraph{Solution: Global Circuit Scheduling}
To correctly prioritize circuits, KIST modifies the way Tor responds to the
libevent writable notifications. Rather than immediately scheduling circuits and
writing cells, KIST instead adds the writable socket to a set of pending
sockets. KIST continues collecting this set over a configurable time period in
order to improve priority by increasing the number of candidate circuits whose
cells may be written. At the end of the period, KIST chooses from the set of all
circuits that contain cells that are waiting to be written to one of the sockets
in the set of pending sockets. We present the details of our implementation of
this approach is \S\ref{sec:deploy}.

\paragraph{Problem: Bloated Socket Buffers}
Jansen \etal observed that relays may have several thousands of TCP sockets
opened at any time, and that the size of each of their send buffers are
automatically tuned (monotonically increased) by the kernel in order to ensure
that the connection can meet the bandwidth delay product and fully utilize the
link. TCP-autotuning~\cite{weigle2002comparison} increases throughput when few
sockets are active, but was found to cause bufferbloat and increase kernel
queuing time in Tor networks where hundreds or
thousands of sockets may be simultaneously active.

\paragraph{Solution: Socket Write Limits}
To reduce bufferbloat, KIST limits how much it will write to each TCP socket
based on the TCP connection's current congestion window and the number of
unacknowledged packets. These values are collected from the kernel using
\texttt{getsockopt(2)} on level \texttt{SOL\_TCP} for option \texttt{TCP\_INFO},
and used to estimate the amount of data that the kernel could
immediately send out into the network (i.e., it would not be throttled by TCP).
KIST limits how much it will write to a socket by the minimum of this estimate
and the amount of free space in each socket buffer.
Finally, KIST includes a global write limit across all sockets to ensure that
the amount of data written to the kernel is not more than the network interface would be able to send.
\section{KIST Implementation} \label{sec:deploy}

We now describe our implementation of KIST to support both global circuit
scheduling and socket write limits. We highlight optimizations that we made to
the original algorithm in order to make KIST more suitable for a production
environment.

\subsection{Supporting Global Circuit Scheduling}

After discussing KIST with the Tor developers, they refactored the
 socket writing logic described in \S\ref{sec:background:traffic} into a new
\textit{socket scheduler} that manages the process of writing cells from the
circuits to the output buffers and the kernel (Figures~\ref{fig:buffers}\circled{E}
and \ref{fig:buffers}\circled{F}). The Tor developers also implemented a
\textit{socket scheduling policy}\footnote{Both the socket scheduler and the default
policy were merged into Tor version \texttt{0.2.6.2-alpha} in December 2014.} that
(i) runs the socket scheduler to write cells to the kernel immediately whenever a circuit has pending cells, and
(ii) follows Tor's previous default behavior of writing as much as possible from pending circuits to writable sockets.
We call this the ``as much as possible" (AMAP) socket scheduling policy.

Although AMAP maintains Tor's previous functionality, it also inherits its
limitations. In particular, because AMAP writes as much as possible and as often
as possible, libevent essentially dictates that sockets get written in a
non-configurable order and therefore circuit priority is ineffective.  However, the new
scheduling framework allows us to fully implement KIST: it allows for the
queuing of writable sockets and for delaying the process of writing to those
sockets.

\subsection{Supporting Socket Write Limits}

We refactored Tor socket scheduling code in order to allow for the implementation of
multiple, distinct socket scheduling policies, and implemented KIST to
limit the amount of data that is written to the kernel.

\subsubsection{KIST Implementation}

\begin{algorithm}[t]
\begin{algorithmic}[1]
\STATE $L_{s} \gets getPendingSockets()$
\FOR{$s$ \textbf{in} $L_{s}$}
\STATE $s.updateTCPInfo()$
\ENDFOR
\WHILE{$len(L_{s}) > 0$}
\STATE $s \gets priorityPop(L_{s})$
\STATE $s.circSched.flush(1)$
\STATE $s.writeOutbufToKernel()$
\IF{$s.circSched.hasCells()$ \AND $s.canWrite()$}
\STATE $priorityPush(L_{s}, s)$ \COMMENT{also updates priority}
\ENDIF
\ENDWHILE
\end{algorithmic}
\caption{\small The KIST socket scheduling policy.}
\label{alg:kistsched}
\end{algorithm}

Algorithm~\ref{alg:kistsched} presents KIST, which
we implemented in Tor \texttt{0.2.8.10} and are preparing to submit to the Tor developers for
merging into a future release. When using KIST, the socket scheduler is executed on a configurable repeating period.

First, KIST performs one system call for each pending
socket in order to update its cache of TCP information. In
\S\ref{sec:neteval:overhead} we evaluate the overhead of this process. Second,
KIST chooses the socket to which it will write using the $priorityPop$ function,
which returns the pending socket with the best priority circuit. KIST writes one
cell from this circuit to Tor's outbuf and immediately flushes it to the kernel in order to
maintain inter-socket priority and avoid the non-deterministic flush order that
is normally imposed by libevent.\footnote{For performance reasons, KIST actually flushes a just-written output buffer to the kernel only when (i) the next scheduling decision would cause a write to a new socket, or (ii) no pending sockets remain.}

KIST then uses $s.canWrite$ to check both that the socket can be written to and
that the write amount has not reached the per-socket limit defined by \newline
\centerline{$limit \gets (2 \cdot cwnd \cdot mss) - (una \cdot mss) - notsent$} \newline
where $cwnd$ is size of the congestion window in packets, $una$ is the number of
sent but unacked packets, $mss$ is the maximum segment size, and $notsent$ is
the number of bytes written to the kernel that have not yet been sent.\footnote{The Linux kernel provides $notsent$ when \texttt{TCP\_INFO} is queried as of version 4.6 (released 2016-05-15); on older kernels, it can be retrieved using \texttt{ioctl(2)} with request \texttt{SIOCOUTQNSD}.} This
slight variation on the previously proposed per-socket
limit~\cite{jansen2014kist} ensures that the kernel can immediately send packets in response to incoming acks rather than waiting for Tor to write more data
the next time that the scheduler is run. If the socket can be written to and the
limit has not been reached, $priorityPush$ returns the socket to the pending
list with its updated priority.

\subsubsection{KIST Optimizations}

During our implementation and testing of KIST, we made the following observations
that led us to reduce its overhead and complexity.

\paragraph{Ignore Idle Sockets}
Jansen \etal~\cite{jansen2014kist} suggested that Tor collect information on
every open socket that is connected to another Tor relay. The
Tor network consists of approximately 7,500 relays \cite{metrics}, which
serves as an upper bound on the number of sockets that may need scheduling. Our
live network tests reveal a fast relay can expect to be connected to 3,000-4,000
others at any time. However, as our overhead analysis in
\S\ref{sec:neteval:overhead} shows, our Tor relay never accumulated more than
127 \textit{pending} sockets (those with available write-pending cells) in a 10 millisecond period.
By only updating TCP information on these
pending sockets, we can greatly reduce the amount of time spent
making system calls.

\paragraph{Ignore Socket Buffer Space}
Jansen \etal~\cite{jansen2014kist} suggested a per-socket write
limit of the minimum between free socket buffer space and TCP's congestion
window. However, we found that the congestion window was the limiting
factor is the vast majority of cases. We can reduce the number of system calls
per socket from three to one by ignoring socket buffer space entirely.
Even if the socket buffer were to run out of space, we can expect that the
kernel will push back and propagate the socket's non-writable state to libevent,
which will prevent Tor from attempting to write to it.

\paragraph{Ignore Global Write Limit}
Jansen \etal~\cite{jansen2014kist} suggested that KIST should enforce a global
write limit across all sockets (in addition to per-socket write limits). We did
not implement this enforcement in order to reduce code complexity, since our
large network simulations described in \S\ref{sec:simeval} show that a global
write limit is unnecessary for preventing bufferbloat given that the per-socket
limits are in place.


\vspace{-1mm}
\section{Simulation Evaluation} \label{sec:simeval}

In this section, we describe our Tor network evaluation of KIST and show its
performance impact across a variety of network conditions.

\vspace{-1mm}
\subsection{Private Tor Network} \label{sec:simeval:net}

We evaluate KIST using
Shadow~\cite{shadowdev,jansen2012shadow}, a discrete-event network simulation
framework. Shadow uses function interposition to intercept all necessary system
calls and redirect them to their simulated counterpart, thereby emulating a
Linux operating system to any applications it runs. Shadow transparently
supports applications that create threads, open UDP and TCP sockets, read and
write to sockets, perform blocking system calls, etc. Applications are compiled
as position-independent executables and loaded into Shadow as plug-ins at run time, and then
directly executed for each virtual simulation node that is configured to run it.
Shadow's strong support for network-based distributed systems in general and Tor
in particular make it ideal for evaluating network-wide effects of new Tor
algorithms.

Shadow directly executes Tor as virtual processes that are connected through a
simulated network. Although a Tor process will attempt to connect to the live, public
Tor network by default, we utilize Tor's private Tor network configuration option
to create a network with our own relays, clients, and servers---all hosted within
the Shadow simulation framework and without direct Internet access.

%
%
%

\paragraph{Virtual Hosts}
We generated a private Tor network using the methods of Jansen
\etal~\cite{cset12modeling} and public Tor metrics data from \texttt{2017-01}.
Our base configuration included a total of 2,000 Tor relays, 49,800 Tor clients,
and 5,000 file servers. The client behavior is as follows. Each of 300 \textit{ShadowPerf}
clients downloads a 50 KiB, 1 MiB, or 5 MiB file and pauses 60-120 seconds
before repeating the download on a new circuit. This behavior mimics the
\textit{TorPerf} download pattern that is used in the public Tor network to
benchmark performance over time~\cite{metrics}, and allows us to
understand the fidelity to the public Tor network. Each of 1495
\textit{bulk} clients repeatedly download 5 MiB files without pausing, while
each of the 48,005 \textit{web} clients download 320 KiB files and pause for a
time between 1 and 60 seconds (chosen uniformly at random) before downloading
another file.

\paragraph{Internet Model}
Shadow uses a connected graph to represent the Internet paths between virtual
simulation hosts. Vertices in the graph correspond to Internet routers to which
a host can be connected while edges correspond to paths between routers and
contain latency and packet loss attributes that Shadow uses to model the path
characteristics. We use the Internet graph of Jansen
\etal~\cite{jansen2014kist}, a complete graph that specifies latency and packet
loss rates between every pair of vertices. However, we made some modifications
because it did not contain accurate packet loss rates on edges.
We did not find a good source of Internet packet loss rates, and so we created a
model where packet loss corresponds to the latency of an edge. First, we reduced
the maximum latency allowed on a single edge to 300 milliseconds to remove
long-tail outliers. Second, we set packet loss rates on the edges in the
complete graph according to the following linear function of the latency of the
edge (in milliseconds): $packetloss \gets latency / (300)(1.5\%)$.
Note that constructing an updated and more accurate graph for Shadow simulation
is outside the scope of this paper, but is a problem that future work should
consider.

\begin{figure*}[t]
\centering
\subfloat[All Clients]{\label{fig:perf:load:ttfb}{\includegraphics[width=0.33\textwidth]{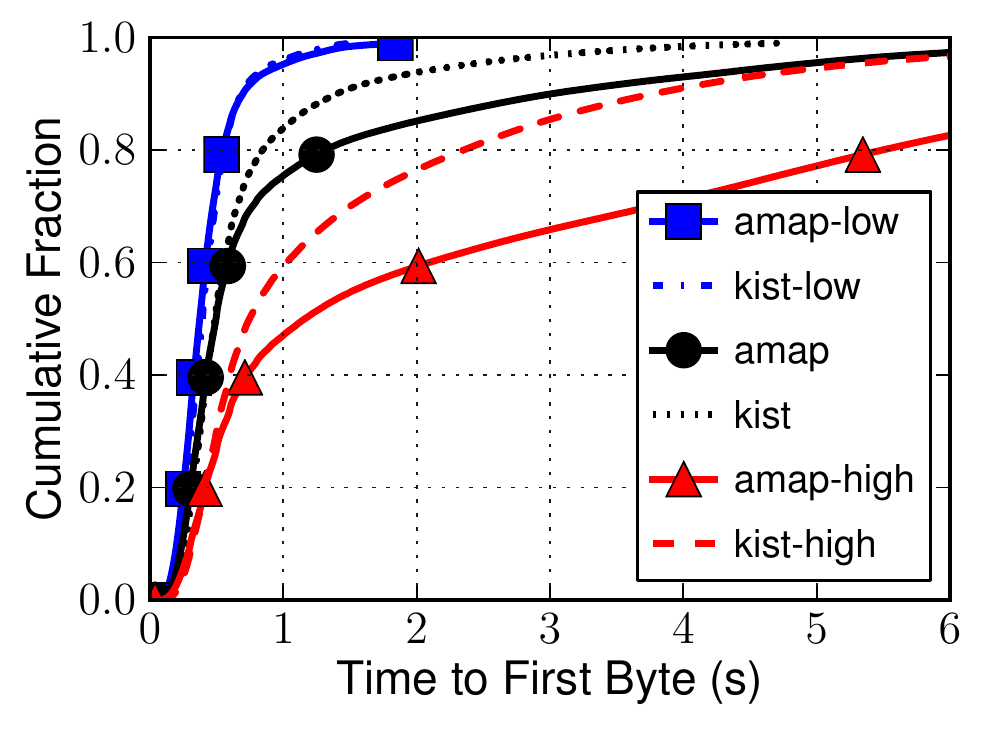}}}
\subfloat[320 KiB Clients]{\label{fig:perf:load:ttlb-web}{\includegraphics[width=0.33\textwidth]{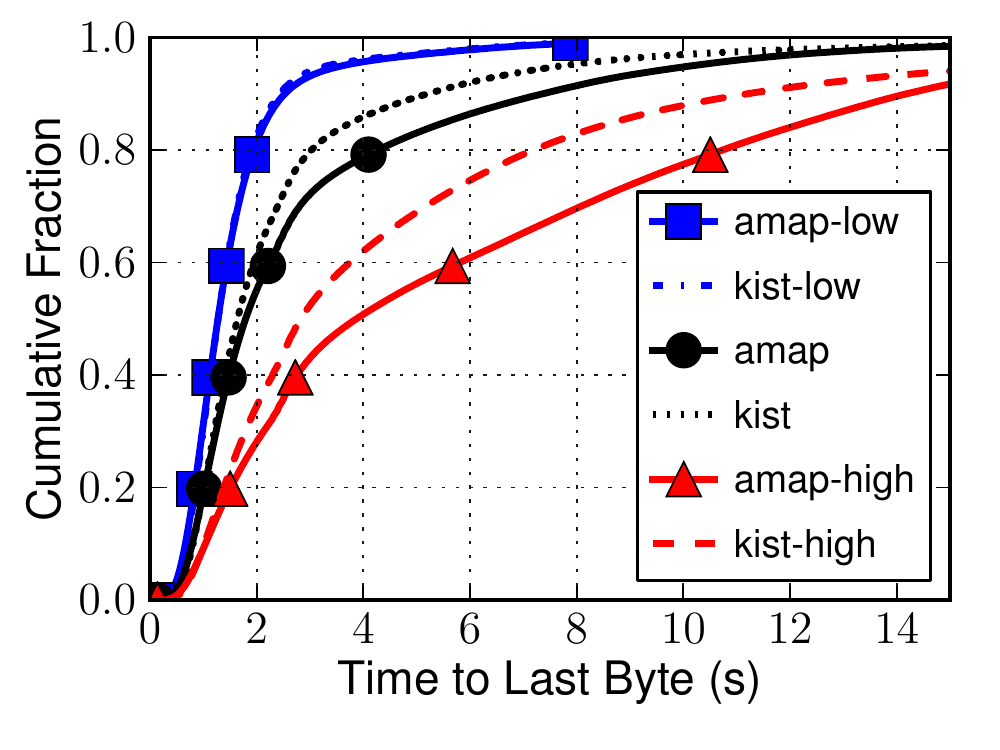}}}
\subfloat[5 MiB Clients]{\label{fig:perf:load:ttlb-bulk}{\includegraphics[width=0.33\textwidth]{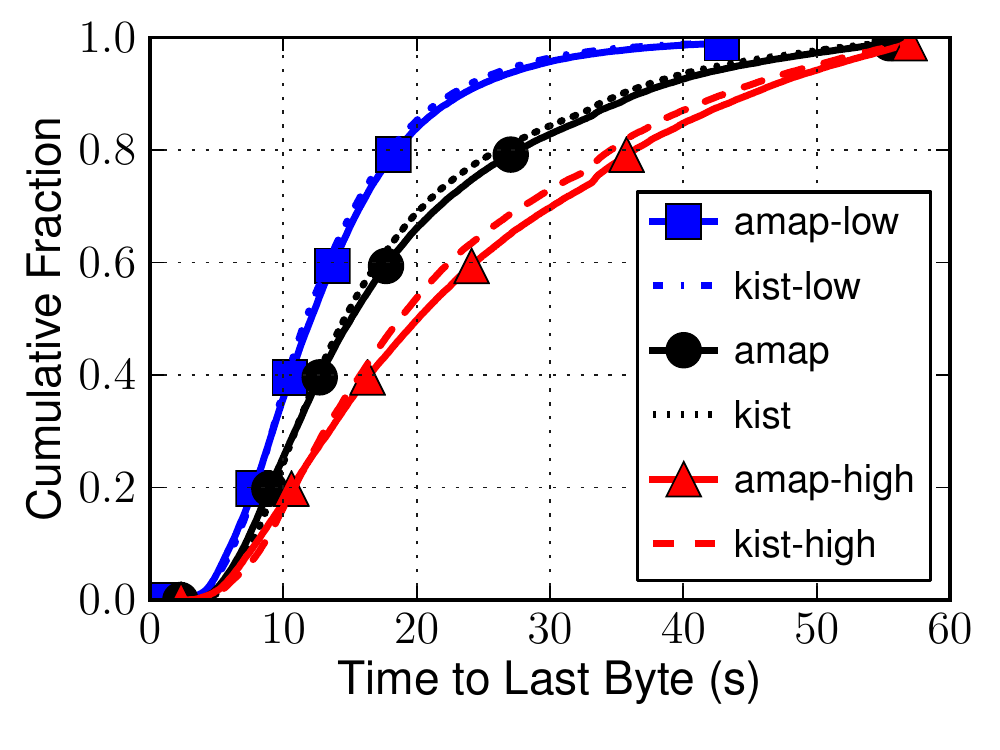}}}
\\
\subfloat[All Relays]{\label{fig:perf:load:tput}{\includegraphics[width=0.33\textwidth]{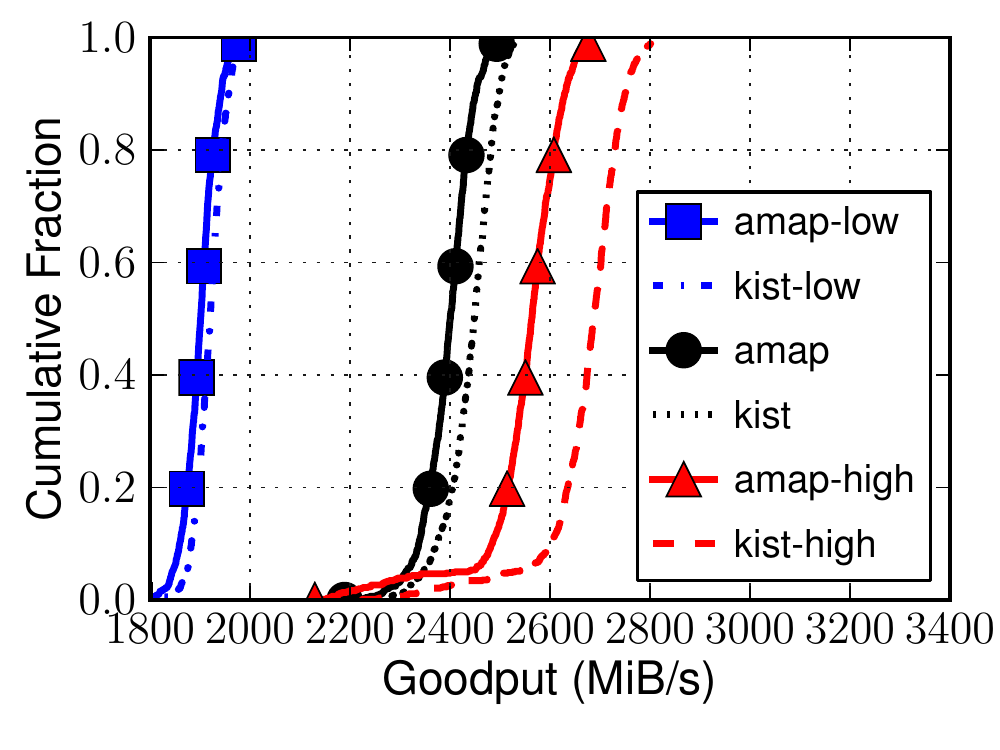}}}
\subfloat[All Relays]{\label{fig:perf:load:kernelqtime}{\includegraphics[width=0.33\textwidth]{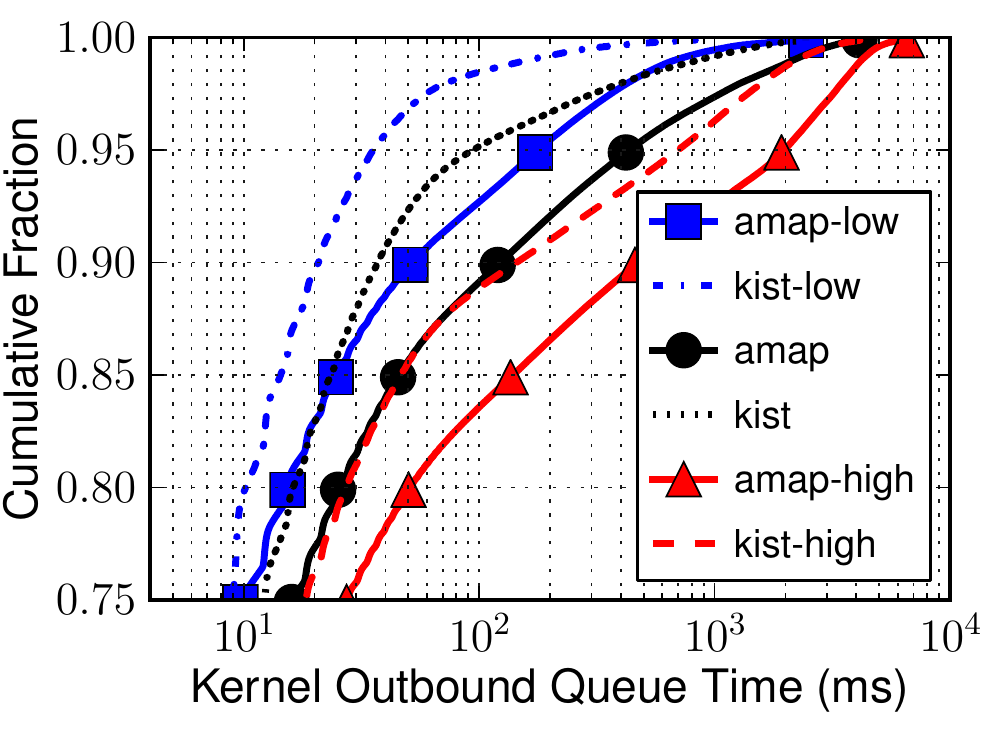}}}
\subfloat[All Relays]{\label{fig:perf:load:torqtime}{\includegraphics[width=0.33\textwidth]{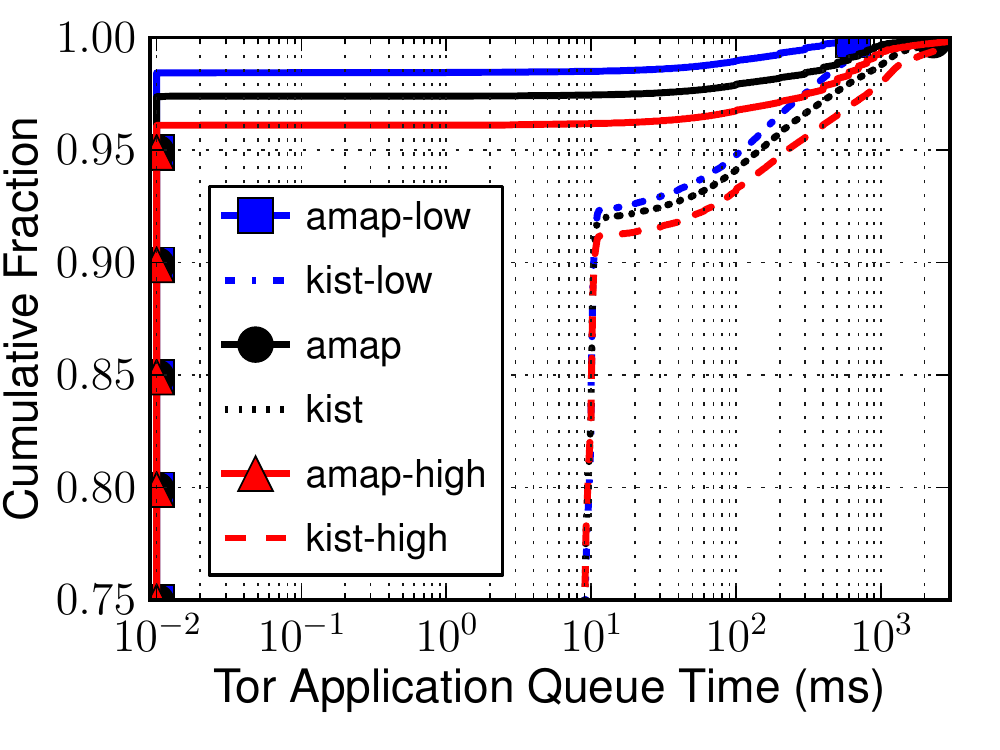}}}
\vspace{-3mm}
\caption{Client and relay performance aggregated across clients and relays for our varying traffic load models.}
\label{fig:perf:load}
\end{figure*}

\vspace{-2mm}
\subsection{Experiments}

Using the private Tor network described above, we ran Shadow experiments with
both the AMAP and the KIST socket scheduling policies that were described in
\S\ref{sec:deploy}. For the KIST experiments, the socket scheduler was
configured to run every 10 milliseconds, as previous work has shown this to
provide good performance~\cite{jansen2014kist}. Additionally, we experimented
with the following variants on the previously described base network.

\paragraph{Traffic load}
We varied the base network to understand how network load affects KIST: we
created \textit{low load} and \textit{high load} network variants by removing
and adding 19,600 clients, respectively. 

\paragraph{Packet loss}
We varied the base Internet model to understand how packet loss rates affect
KIST: we created a \textit{no~loss} model with all packet loss rates set to 0,
and we created a \textit{high~loss} model with packet loss rates double that of
the base model (to a maximum of 3\%).

We ran 10 experiments in total: one for AMAP and one for KIST for the
base network as well as each of the four variants. Each experiment consumed
between 0.5 and 1 TiB of RAM and simulated 45 minutes of full network activity in
6 to 8 days running on our hardware. We collected client download times, and
we instrumented our relays to collect goodput information
(using existing Tor control port mechanisms), Tor cell queuing times (using a
cell tracing patch we developed), and kernel queuing times (with a Shadow patch).

\subsection{Results}

We evaluate and compare KIST and AMAP across a variety of different performance metrics.

\begin{figure*}
\centering
\subfloat[All Clients]{\label{fig:perf:loss:ttfb}{\includegraphics[width=0.33\textwidth]{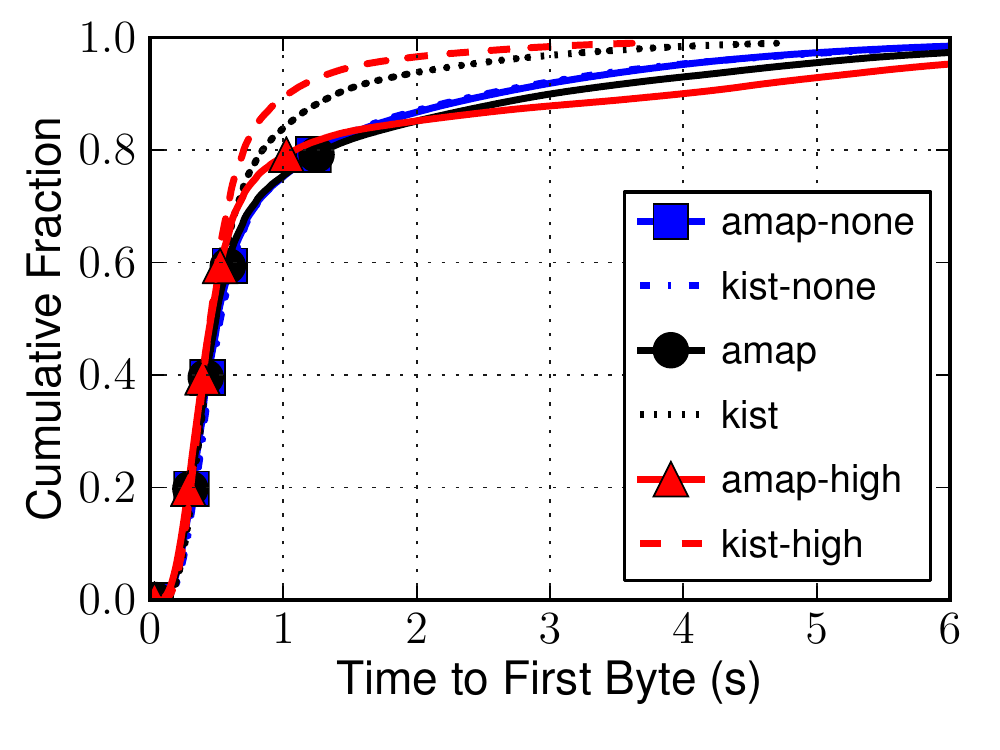}}}
\subfloat[320 KiB Clients]{\label{fig:perf:loss:ttlb-web}{\includegraphics[width=0.33\textwidth]{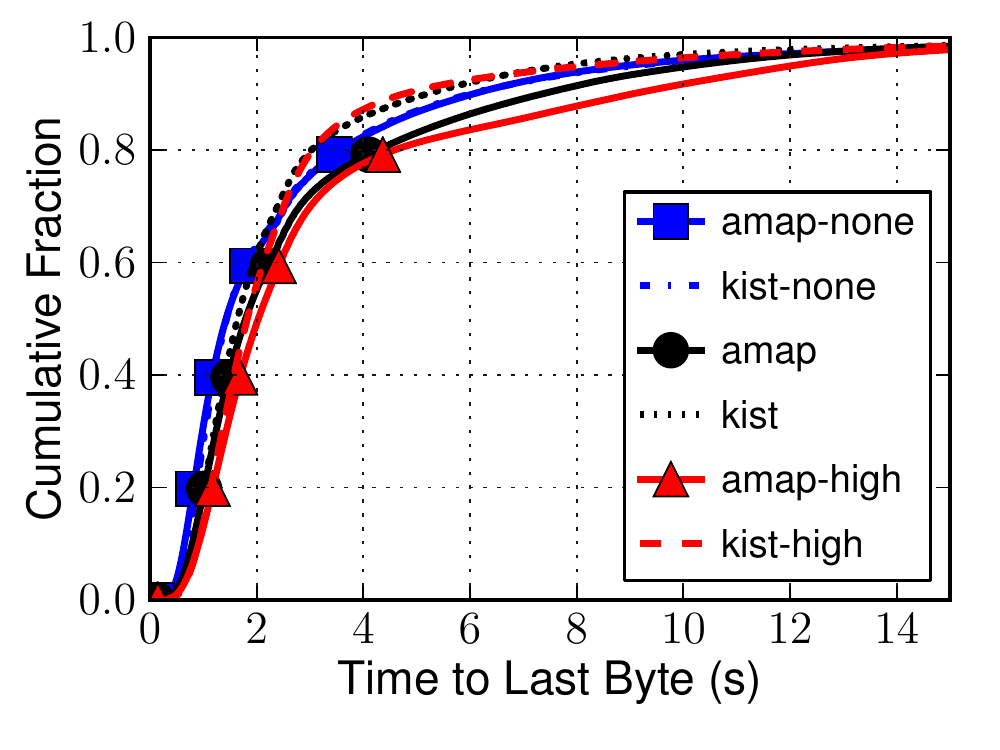}}}
\subfloat[5 MiB Clients]{\label{fig:perf:loss:ttlb-bulk}{\includegraphics[width=0.33\textwidth]{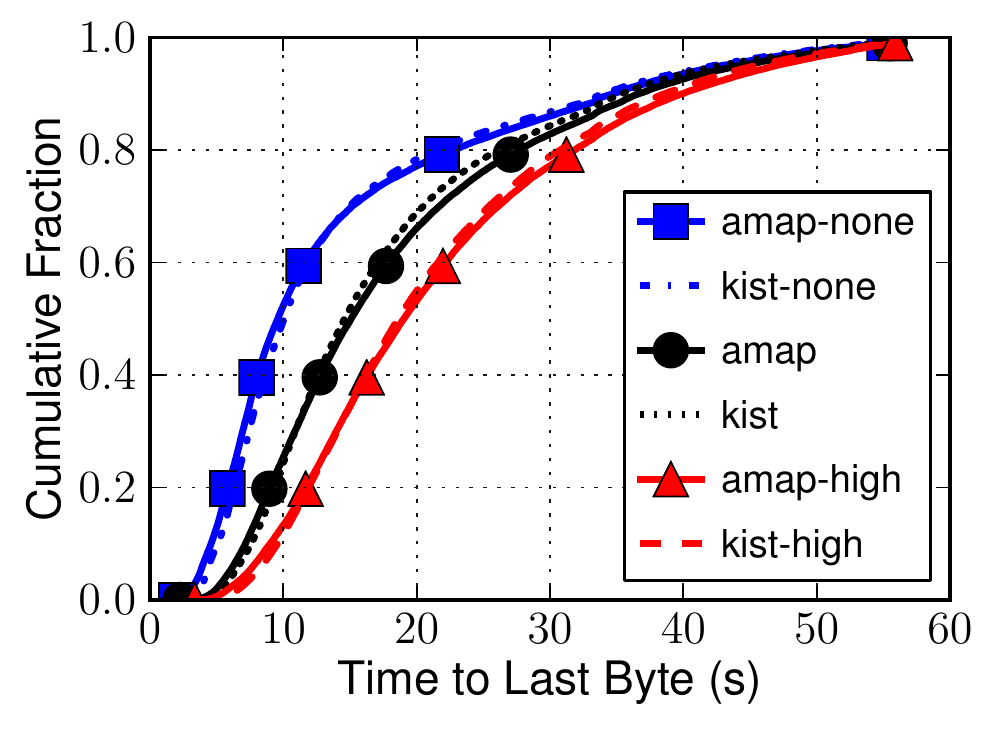}}}
\\
\subfloat[All Relays]{\label{fig:perf:loss:tput}{\includegraphics[width=0.33\textwidth]{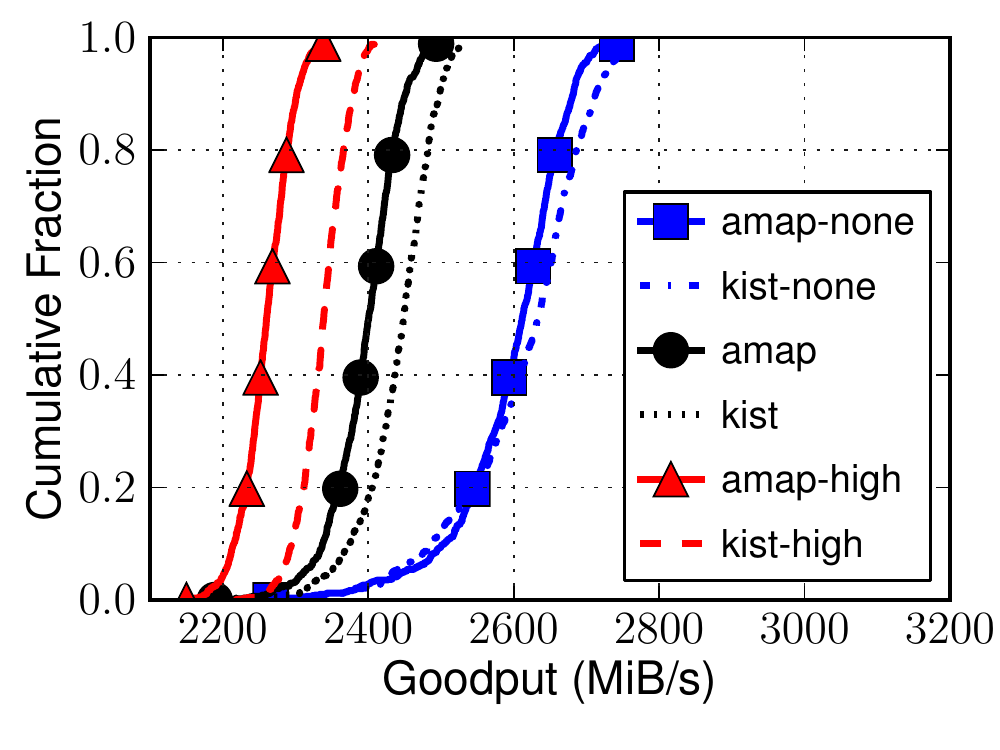}}}
\subfloat[All Relays]{\label{fig:perf:loss:kernelqtime}{\includegraphics[width=0.33\textwidth]{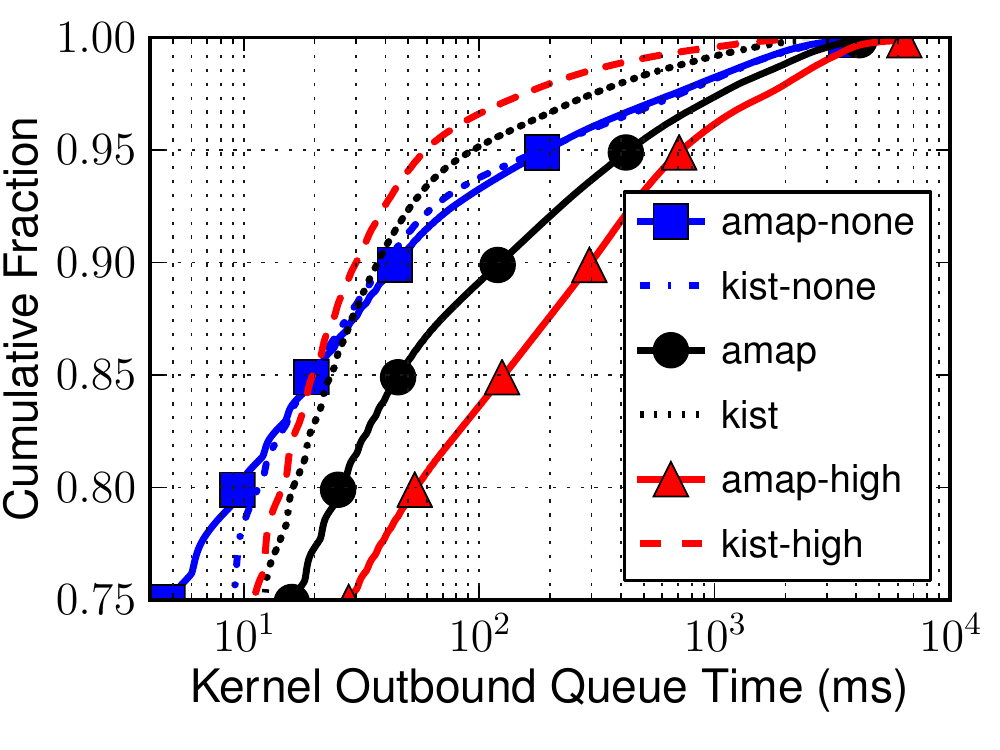}}}
\subfloat[All Relays]{\label{fig:perf:loss:torqtime}{\includegraphics[width=0.33\textwidth]{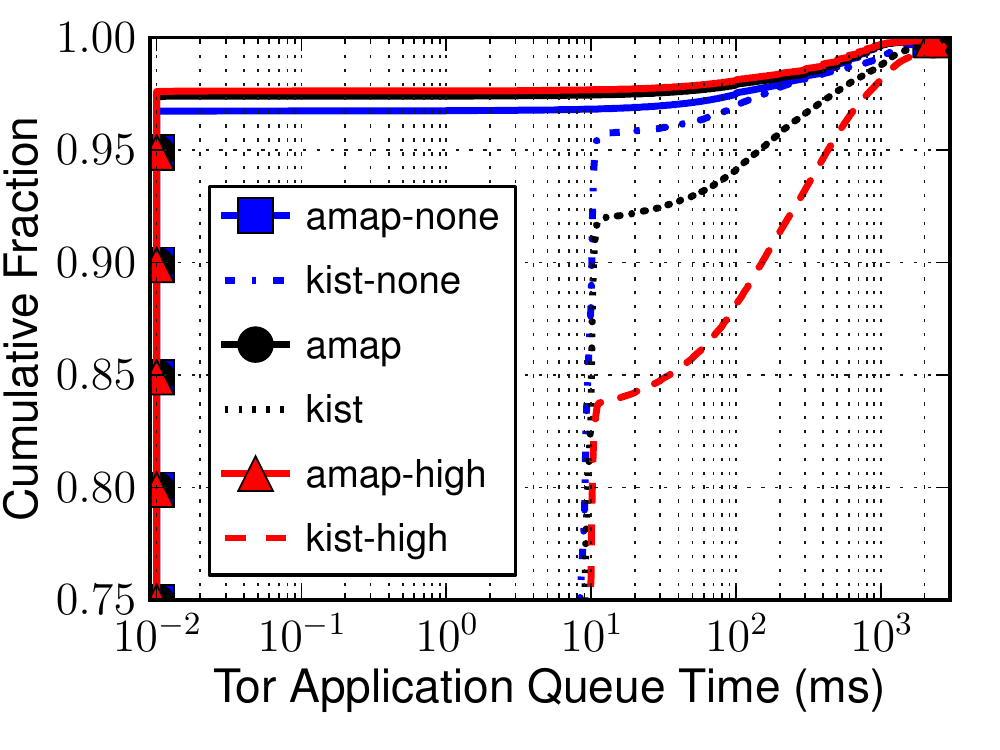}}}
\vspace{-3mm}
\caption{Client and relay performance aggregated across clients and relays for our varying packet loss models.}
\label{fig:perf:loss}
\end{figure*}

\paragraph{Effects of Traffic Load}
The performance effects of traffic load on AMAP (solid lines) and KIST
(non-solid lines) are shown in Figure~\ref{fig:perf:load}, where the line colors
indicate the low, normal, and high load models.

Client performance is shown in
Figures~\ref{fig:perf:load:ttfb}-\ref{fig:perf:load:ttlb-bulk} as the time to
reach the first and last byte for all completed client downloads, across the low,
regular, and high traffic load models. We make three major observations from
these results. First, when there is low traffic load on the network, clients
download times are generally unaffected by the choice of scheduling policy (all
of the blue lines in \ref{fig:perf:load:ttfb}-\ref{fig:perf:load:ttlb-bulk}
showing download times under low packet loss are roughly overlapping).
Second, download times increase across all scheduling
policies as the load on the network increases, but the increase is greater
for AMAP than for KIST (i.e., downloads with KIST finish more quickly than those with
AMAP as load increases). Third, client performance when using KIST
is no worse and generally much better than when using AMAP, but the
improvement over AMAP diminishes as download size increases and the EWMA circuit
scheduler's priority mechanisms become effective at preferring lower-throughput
flows.

Relay performance is shown in
Figures~\ref{fig:perf:load:tput}-\ref{fig:perf:load:torqtime}.
Figure~\ref{fig:perf:load:tput} shows that aggregate Tor network goodput per second is
higher when using both scheduling policies as the network load increases, matching
intuition. Goodput increases over AMAP when using KIST as network load
increases, but the improvement that KIST provides is most significant on the
highest-load model that we tested. Figure~\ref{fig:perf:load:kernelqtime} shows
that KIST generally reduces kernel queue time by more than it increases Tor
queue time as shown in Figure~\ref{fig:perf:load:torqtime}, suggesting that KIST is
capable of reducing congestion overall rather than simply relocating it. Note
that Tor queue time in Figure~\ref{fig:perf:load:torqtime} is nearly zero for AMAP
across all three load models, as Tor writes as much as possible to the kernel
and tends to not queue data in the application layer. Also note that the sharp elbow at
10 milliseconds for KIST is due to the configured interval in which the scheduler is run.

\paragraph{Effects of Packet Loss}
The performance effects of packet loss for AMAP (solid lines) and KIST
(non-solid lines) are shown in Figure~\ref{fig:perf:loss}, where the line colors
indicate the no, normal, and high packet loss models.

Client performance is shown in
Figures~\ref{fig:perf:loss:ttfb}-\ref{fig:perf:loss:ttlb-bulk}. Recall that the
general trend with the varying load models was that KIST and AMAP both reduced
client performance as load increased, but the reduction when using KIST was less
than when using AMAP. However, the general trend when varying packet loss is a bit
different. When no packet loss is present, similar performance is achieved with
both AMAP and KIST. However, as packet loss increases, AMAP tends to worsen
low-volume client performance while KIST tends to improve it.
Figure~\ref{fig:perf:loss:ttfb} and~\ref{fig:perf:loss:ttlb-web} both show that
KIST actually performs best with high packet loss while AMAP performs worst, while
Figure~\ref{fig:perf:loss:ttlb-bulk} shows similar results as for varying load:
the improvement diminishes for 5 MiB downloads since the circuit priority
mechanism is operating effectively.
We suspect that KIST achieves better performance for low-volume traffic at
higher packet loss rates because it effectively deprioritizes high-volume
circuits; the less-congested kernel can then react more quickly to low-volume
traffic as Tor prioritizes its delivery to the kernel. More work is needed to
verify this suspicion.

Relay performance is shown in
Figures~\ref{fig:perf:loss:tput}-\ref{fig:perf:loss:torqtime}.
Figure~\ref{fig:perf:loss:tput} shows that aggregate Tor network goodput
decreases as packet loss increases, but it decreases less when using KIST than
when using AMAP. Figure~\ref{fig:perf:loss:kernelqtime} shows again that KIST is
able to reduce kernel queue time more as packet loss rates increase, while AMAP
increases kernel queue times as packet loss rates increase. Finally, the general
trends in Figure~\ref{fig:perf:loss:torqtime} are similar that of the Tor queue
times under the varying load models: Tor queue time is nearly zero when using AMAP for
all tested loss models, while it is 10 milliseconds or less for over 80 percent of the data when
using KIST.


\section{Live Network Evaluation} \label{sec:neteval}

In this section, we evaluate KIST by running it on a 
live relay in the public Tor network.

\subsection{Deploying KIST}

The KIST scheduling decisions are all local to each relay, and function
independent of the other relays in a circuit. As a result, KIST is
naturally incrementally deployable. This allows us to deploy KIST on a single
relay under our control in the public Tor network to further understand its
real-world performance effects.


\begin{figure*}[t]
\centering
\subfloat[All Clients]{\label{fig:perf:live:ttfb}{\includegraphics[width=0.33\textwidth]{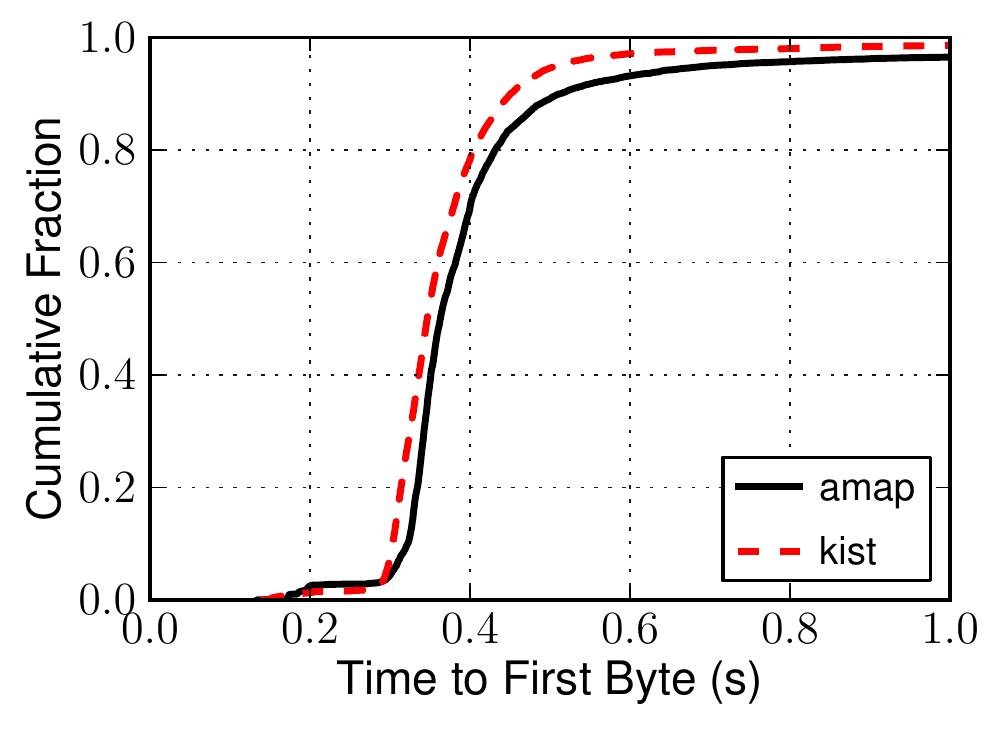}}}
\subfloat[320 KiB Clients]{\label{fig:perf:live:ttlb-web}{\includegraphics[width=0.33\textwidth]{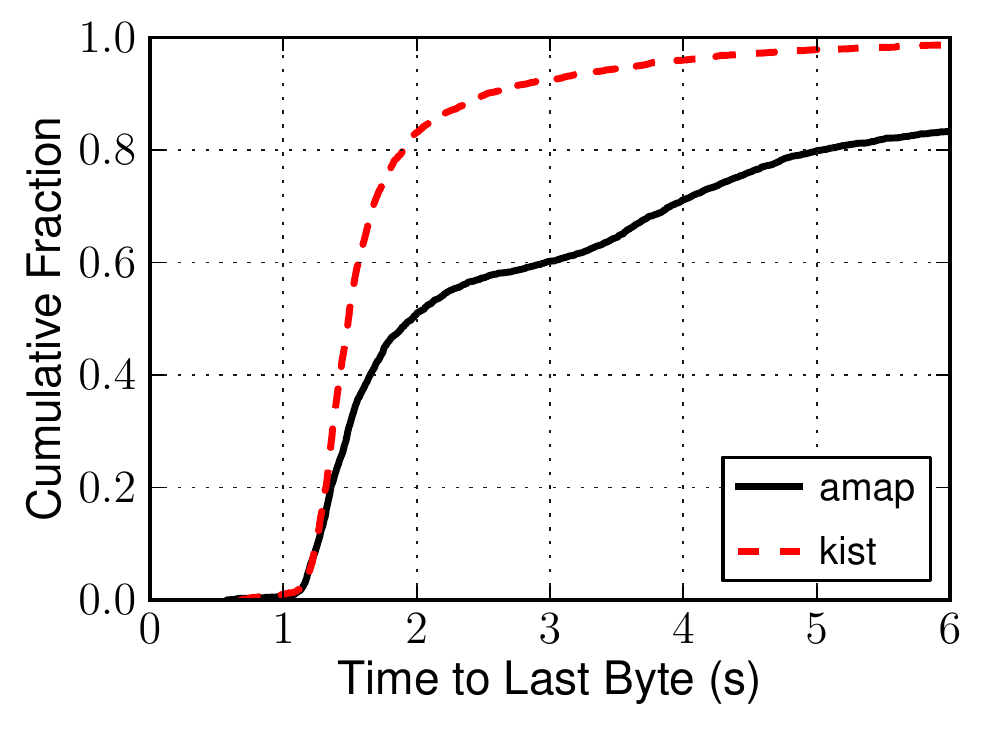}}}
\subfloat[5 MiB Clients]{\label{fig:perf:live:ttlb-bulk}{\includegraphics[width=0.33\textwidth]{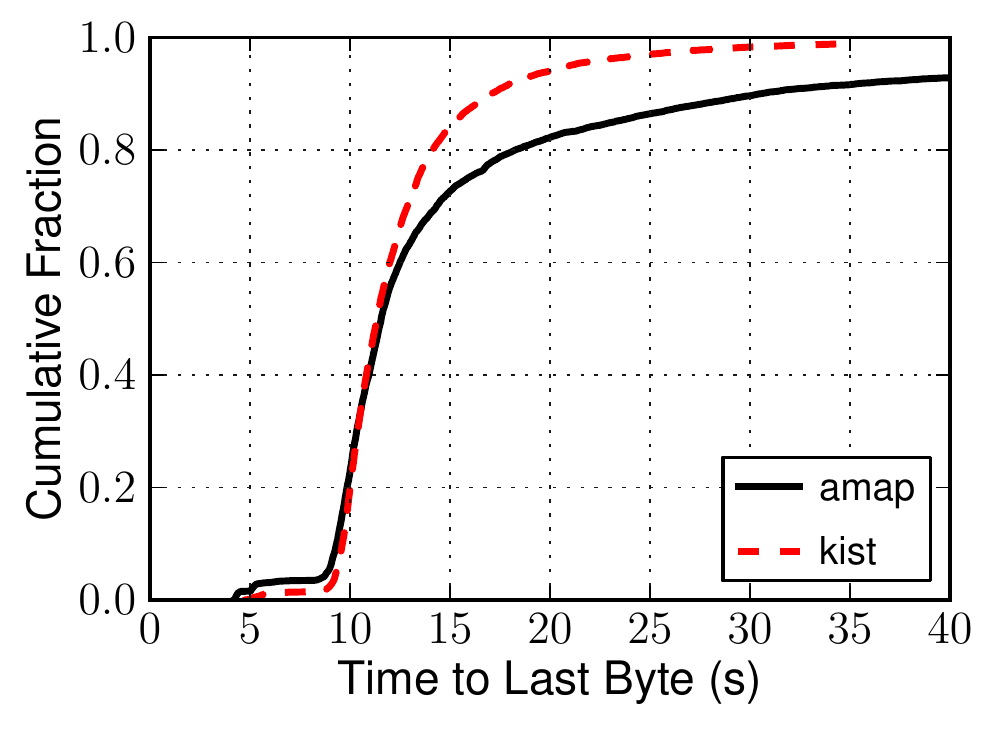}}}
\vspace{-3mm}
\caption{Client performance aggregated across five clients downloading through the live Tor network.}
\label{fig:perf:live:client}
\end{figure*}

\begin{figure*}[t]
\centering
\subfloat[]{\label{fig:perf:live:tput}{\includegraphics[width=0.33\textwidth]{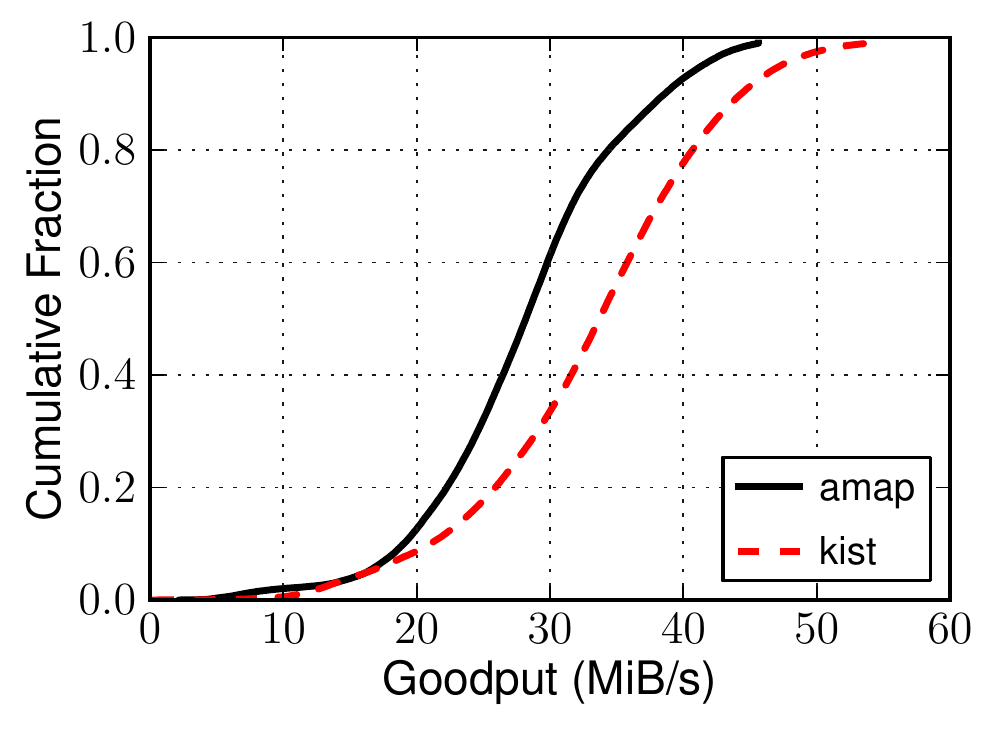}}}
\subfloat[]{\label{fig:perf:live:kernelqtime}{\includegraphics[width=0.33\textwidth]{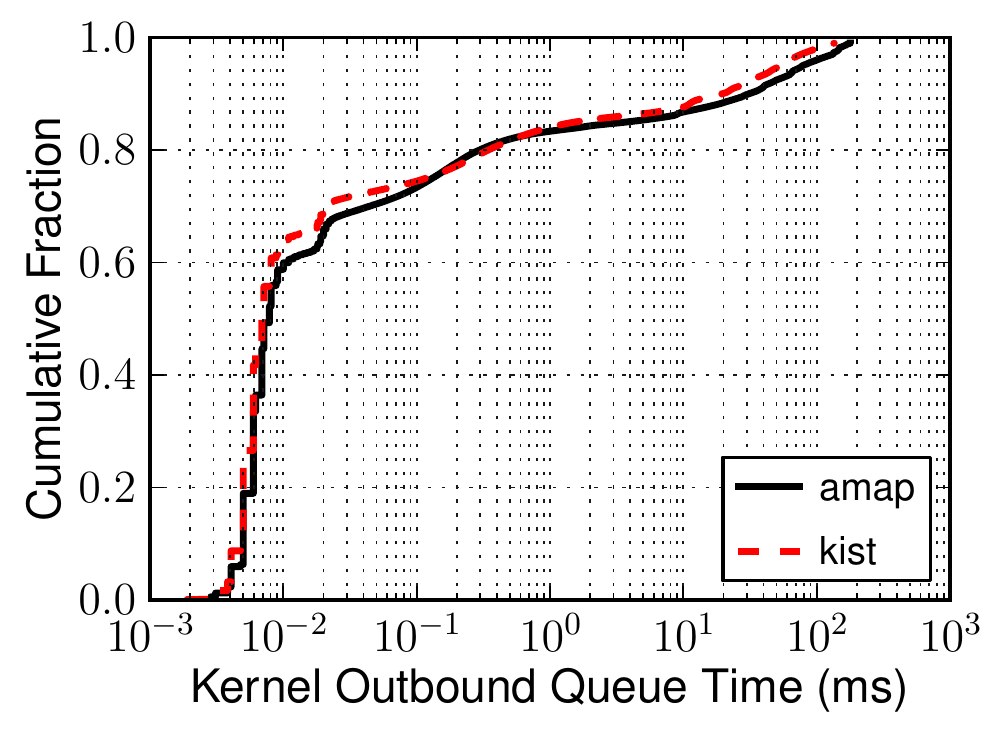}}}
\subfloat[]{\label{fig:perf:live:torqtime}{\includegraphics[width=0.33\textwidth]{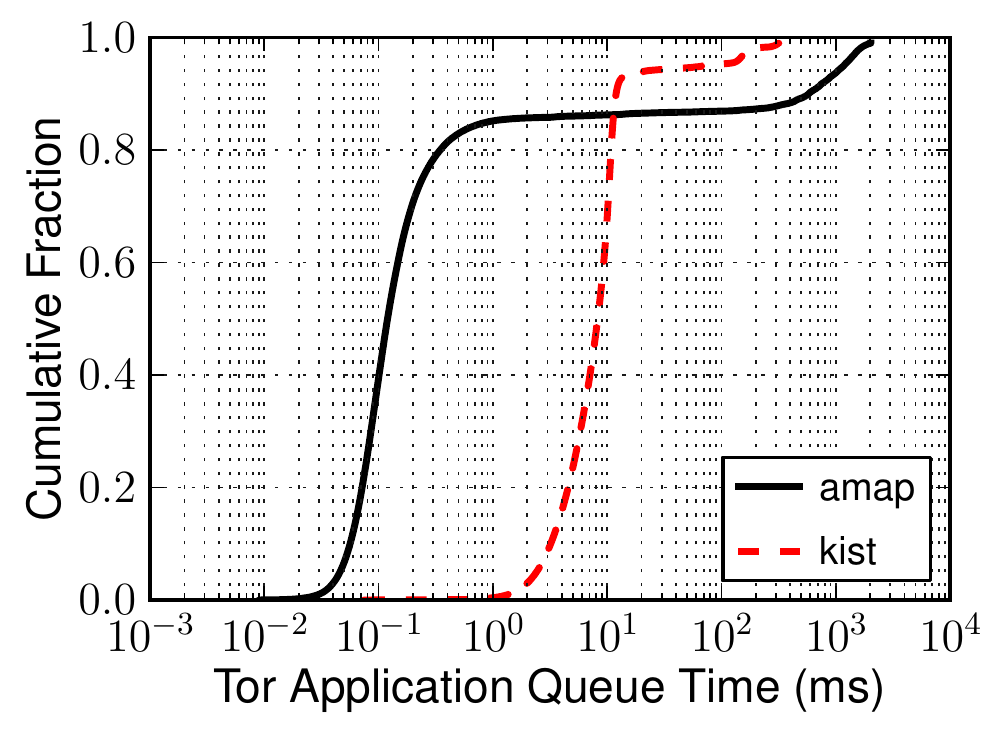}}}
\vspace{-3mm}
\caption{Tor relay performance on our fast exit relay running in the live Tor network.}
\label{fig:perf:live:relay}
\end{figure*}

We ran a Tor exit relay\footnote{The relay fingerprint was
\texttt{0xBCCB362660}.} with the default exit policy on a bare-metal machine
rented from Hurricane Electric (an Internet service provider). The machine had a
4-core/8-thread intel Xeon E3-1230 v5 CPU running at 3.40 GHz, and was connected
to a unmetered access link capable of a symmetric bandwidth of 1 Gbit/s (2
Gbit/s combined transmit and receive). Several unrelated relays were co-hosted
on the machine, but the average combined daily bandwidth
used did not exceed 1.5 Gbit/s.

We ran our relay for several weeks before starting a 2 day experiment where we
ran KIST and AMAP for one day each. During this experiment, we also ran three
\textit{web} clients that download 320 KiB files and pause an average of 60
seconds between downloads, and two \textit{bulk} clients that download 5 MiB
files and pause for 1 second between downloads. The clients choose new entry and
middle relays for every circuit, but we pin our relay as the exit. As in the
Shadow simulations in \S\ref{sec:simeval}, we instrumented our relay to collect
goodput, Tor cell queuing times, and kernel queuing times, and we instrumented
our clients to collect file download times.



\subsection{Results} \label{sec:neteval:overhead}


During our experiment, the web clients finished 7,770 downloads and the bulk
clients finished 18,989 downloads. Figure ~\ref{fig:perf:live:client} shows the
distributions of download times recorded by our clients. While KIST reduced
download times relative to AMAP across all metrics, the relative improvement was
greater for 320 KiB files (Figure~\ref{fig:perf:live:ttlb-web}) than for 5 MiB
files (Figure~\ref{fig:perf:live:ttlb-bulk}); this could indicate that circuit
priority was more effective under KIST, although we note that there may be
network effects outside of our control that are also influencing the results.

Figure~\ref{fig:perf:live:relay} shows the performance results collected on our relay.
Figure~\ref{fig:perf:live:tput} shows that KIST increased goodput relative to
AMAP during our observational period. Although
Figure~\ref{fig:perf:live:kernelqtime} shows an insignificant change in kernel
outbound queue time, Figure~\ref{fig:perf:live:torqtime} shows that KIST
increased Tor application queuing time by less than 10 milliseconds (the
configured scheduling interval) for over 90 percent of the sampled cells; we
suspect that the relatively higher Tor queue time for the remaining sampled
cells is due to the circuit scheduler effectively de-prioritizing high-volume
circuits. Additionally, KIST reduced the worst case application queue times from
over 2,000 milliseconds to less than 400 milliseconds.


We also collected the overhead of performing the \texttt{getsockopt(2)} call to
retrieve TCP information for write-pending sockets. We observed that the median
number of write-pending sockets that accumulated during a 10 millisecond period
was 23 (with min$=$1, q1$=$18, q3$=$27, and max$=$127), while the median amount
of time to collect TCP information on all write-pending sockets was 23
microseconds (with min$=$1, q1$=$17, q3$=$33, and max$=$674). We observed a
linear relationship between the amount of time required to collect TCP
information on all write-pending sockets and the number of such sockets (1.08
microseconds per pending socket), independent of the total number of open sockets.
Therefore, we believe that the KIST overhead, with our
optimization of only collecting TCP information on pending sockets, should be
tolerable to run in the main thread for even the fastest Tor relay.

\section{Conclusion} \label{sec:conc}

In this work, we implemented KIST with a goal of deploying it into the Tor
network. We evaluated its performance impact in simulation under a range of
network load and packet loss conditions, and found that KIST can improve client
and relay performance, particularly when a relay is under high load or high
packet loss. We also ran KIST in the public Tor network, and found that KIST has
an indistinguishable effect on relay throughput and kernel queuing times.
We will release our implementation as open-source software so that it can
be included in a future Tor release.

\paragraph{Lessons Learned}
As with most Tor research, we found it useful to communicate with the Tor
developers early and often. They are experts in developing and maintaining
anonymous communication systems and their collaboration and feedback greatly
improved the quality of this work. Still, we found it to be extremely
time-consuming to produce a working implementation given the complexities of
both the Tor software and the way that it inter-operates with TCP and the
kernel. We advise those interested in deploying Tor research to carefully
compare the costs and benefits of creating new knowledge through additional
research with those of deploying previous research proposals.

\paragraph{Future Work}
Future work should consider creating updated data-driven models of latency and
packet loss rates between relays that would be useful to Tor experimentation
tools like Shadow. This could be done using direct measurement with RIPE Atlas
probes 
or tools like Ting~\cite{imc2015ting}. More work is also needed to verify our
simulation findings that KIST is capable of increasingly improving performance
for low-volume traffic under high load and packet loss rates in the public Tor
network.

\section*{Acknowledgments} 

We 
thank Andrea Shepard for implementing the socket scheduling code, and David
Goulet, Nick Mathewson, and Tim Wilson-Brown for their assistance and feedback
during development.

This work has been partially supported by the Defense Advanced Research Project
Agency (DARPA), the National Science Foundation (NSF) under grant number
CNS-1527401, and the Department of Homeland Security (DHS) Science and
Technology Directorate, Homeland Security Advanced Research Projects Agency,
Cyber Security Division under agreement number FTCY1500057. The views expressed
in this work are strictly those of the authors and do not necessarily reflect
the official policy or position of DARPA, NSF, DHS, or the U.S. Naval Research
Laboratory.

\balance
{\footnotesize 
\bibliographystyle{acm}
\bibliography{ms}
}
 


\end{document}